\documentclass[aps,twocolumn,twoside,nofootinbib,prd,preprintnumbers,superscriptaddress]{revtex4}
\usepackage{graphicx}
\usepackage{amsmath}
\usepackage{amssymb}

\newcommand{\lsim}{\lesssim}
\newcommand{\gsim}{\gtrsim}

\newcommand{\Mpl}{M_{\rm Pl}}

\newcommand{\ra}{\rightarrow}

\begin{document}

%\topmargin 0.25cm

% You should use BibTeX and revtex.bst for references
\bibliographystyle{revtex}

\preprint{ANL-HEP-PR-07-39}

\title{Four Generations and Higgs Physics}

\author{Graham D. Kribs}
\affiliation{\mbox{Department of Physics and Institute of Theoretical Science,
University of Oregon, Eugene, OR 97403, USA}}
 
\author{Tilman Plehn}
\affiliation{\mbox{SUPA, School of Physics, University of Edinburgh, 
Edinburgh EH9 3JZ, Scotland}}

\author{Michael Spannowsky}
\affiliation{\mbox{ASC, Department f\"ur Physik, Ludwig-Maximilians-Universit\"at M\"unchen, 80333 M\"unchen, Germany}}

\author{Tim M.P. Tait}
\affiliation{\mbox{HEP Division, Argonne National Laboratory, 
9700 Cass Ave., Argonne, IL 60439, USA}}

%\date{\today}

\begin{abstract}

In the light of the LHC, 
we revisit the implications of a fourth generation of chiral matter.  
We identify a specific ensemble of particle masses and mixings that
are in agreement with all current experimental bounds as well as 
minimize the contributions to electroweak precision observables.
Higgs masses between 115-315 (115-750)~GeV 
are allowed by electroweak precision data at the $68\%$ and $95\%$ CL.
Within this parameter space, 
there are dramatic effects on Higgs phenomenology:
production rates are enhanced, weak--boson--fusion channels are suppressed,
angular distributions are modified, 
and Higgs pairs can we observed.  We also identify 
exotic signals, such as Higgs decay to same-sign dileptons.
Finally, we estimate the upper bound on the cutoff scale 
from vacuum stability and triviality.

\end{abstract}

% \pacs{}

%\maketitle must follow title, authors, abstract and \pacs
\maketitle

\section{Introduction}
\label{intro-sec}

New physics that affects the observability of the 
Higgs boson of the Standard Model (SM) is of utmost importance to 
study.  One the simplest kinds of new physics is a sequential 
replication of the 
three generations of chiral matter~\cite{footnote0}.  
Such a fourth generation has been considered and forgotten or discarded 
many times, wrongly leaving the impression that it is either ruled out 
or highly disfavored by experimental data 
(for instance, see Ref.~\cite{Yao:2006px}).

The status of four generations is more subtle~\cite{Frampton:1999xi}.
Ref.~\cite{He:2001tp} analyzed the contributions of one (and more)
extra generations to the oblique parameters and explicitly found 
that one generation can be perfectly consistent with a heavy ($500$~GeV) Higgs.
These significant results are primarily based on
numerical scans, with emphasis on the role of a lighter neutrino ($50$~GeV) 
to minimize the contributions to the oblique parameters
(see also Ref.~\cite{Maltoni:1999ta}).
However,
a neutrino with mass of $50$ GeV, if unstable, is ruled out by LEP II bounds,
while if it exactly stable, may be ruled out by dark matter
direct search experiments \cite{footnote1}.
Correlations of the mass parameters leading to viable spectra 
are certainly not transparent, making it hard to determine
how to parse their results against present experimental bounds.

Subsequent analyses~\cite{Novikov:2001md,Novikov:2002tk}
studied the relationships among fourth generation 
parameters, but their analysis was performed using a global (numerical) 
fit to 2001 electroweak data and again emphasized a 50~GeV neutrino.
Electroweak data has since been refined 
(in particularly $M_W$), so these results 
no longer obviously apply, in particular if we incorporate a 
heavier neutrino.  The impact of a chiral fourth generation 
on Higgs physics has been briefly
discussed~\cite{Arik:2001iw,Arik:2002ci,Arik:2005ed}, 
however the range of masses that were considered were not necessarily 
correlated to the fourth generation mass spectra and Higgs mass
appropriate to satisfy current direct search bounds and electroweak
precision constraints.  Moreover, in cases where there is overlap,
our results do not always agree; we point out the differences below.

In this paper, we first systematically determine the allowed parameter 
space of fourth generation masses and mixings.  We find quite simple mass 
relations that minimize the precision electroweak oblique parameters,
so our analysis can easily be extended to future refinements
in electroweak measurements.
We then use typical spectra to compute the consequences for fourth generation 
particle production and decay, as well as the effects on 
the Higgs sector of the Standard Model. We find that a wide range of 
Higgs masses is consistent with electroweak data, 
leading to significant modifications 
of Higgs production and decay.  We outline the major effects, 
identifying the well-known effects from others that (to our knowledge) 
are new.  

There are in addition spectacular signals of the fourth generation itself.
Given that direct searches at LEP II and Tevatron have already
constrained the masses somewhat, we can expect future searches at 
Tevatron will continue to push the limits up, but will not rule out 
four generations.  
The LHC is able probe heavy quarks throughout their mass range.
Many of the signals have been recently considered 
(albeit in somewhat different mass ranges and context from what
we consider here) in Refs.~\cite{Holdom:2006mr,Skiba:2007fw},
to which we refer the interested reader.

\section{Four Generations}
\label{fourgen-sec}

The framework we consider is to enlarge the Standard Model 
to include a complete sequential fourth generation of chiral matter 
($Q_4$, $u_4$, $d_4$, $L_4$, $e_4$) as well as a single right-handed 
neutrino $\nu_4$.   
Yukawa couplings and right-handed neutrino masses are given by
\begin{alignat}{5}
{\cal L} \; =& \;  y^u_{pq} \overline{Q}_p H u_q 
             + y^d_{pq} \overline{Q}_p H^\dagger d_q
             + y^e_{pq} \overline{L}_p H^\dagger e_q   \notag \\
          &{} + y^{\nu}_{pq} \overline{L}_p H \nu_q 
             + \frac{1}{2} M_{pq} \overline{\nu}^c_p \nu_q + {\rm h.c.} \; .
\end{alignat}
The generation indices are $p,q=1,2,3,4$ while we reserve $i,j=1,2,3$
for the Standard Model. SU(2) contractions are implicit.
Light neutrino masses can arise from either a hierarchy in 
neutrino Yukawa couplings $y^{\nu}_{ij} \ll y_{44}$ or
right-handed neutrino masses $M_{ij} \gg M_{44}$ or some
combination.   (For an amusing combination, see Ref.~\cite{Hill:1989vn}).
We mainly consider two possibilities for the fourth--generation neutrino mass:
purely Dirac ($M_{44} = 0$) and
mixed ($M_{44} \sim y^{\nu}_{44} v$).

There are four obvious restrictions on a fourth generation:
(1) The invisible width of the $Z$;
(2) Direct search bounds;
(3) Generational mixing;
(4) Oblique electroweak effects.
We now discuss them one-by-one.\bigskip

Once a fourth--generation neutrino has a mass $m_\nu \gsim M_Z/2$,
the constraint from the invisible $Z$ width becomes moot.
Assuming non-zero mixings $y_{i4}$ or $M_{i4}$,
the fourth--generation quarks, charged lepton, and neutrino 
decay, and thus there are no cosmological constraints 
from stable matter. 
(We will briefly comment on neutrino dark matter at the end the paper.)

A robust lower bound on fourth--generation masses comes from 
LEP~II.  The bound on unstable charged leptons
is $101$~GeV, while the bound on unstable neutral Dirac neutrinos is 
$(101,102,90)$~GeV for the decay modes $\nu_4 \ra (e,\mu,\tau) + W$.
These limits are weakened only by about 10~GeV 
when the neutrino has a Majorana mass.  
Because the small differences in the bounds between different flavors, 
charged versus neutral leptons, and Majorana versus Dirac mass
do not affect our results, we apply the LEP~II 
bounds as $m_{\nu_4,\ell_4,u_4,d_4} \gtrsim 100$~GeV throughout.

The Tevatron has significantly greater sensitivity for 
fourth--generation quarks~\cite{footnote2}.
The strongest bound is from the CDF search 
for $u_4 \overline{u}_4 \ra q\overline{q} W^+W^-$, 
obtaining the lower bound $m_{u_4} > 258$~GeV to 95\% confidence level (CL).
\cite{cdf-tprime}.
No $b$-tag was used, so there is no dependence on the final--state jet flavor,
and hence this limit applies independent of the CKM elements $V_{u_4 i}$.
There is no analogous limit on the mass of $d_4$.
If $m_{d_4} > m_t + m_W$ and $|V_{t d_4}| \gg |V_{u d_4}|,|V_{c d_4}|$,
then the dramatic $d_4 \overline{d}_4 \ra t\overline{t} WW$ signal
may be confused into the top sample.  If the decay proceeds
through a lighter generation, then the production rate and 
signal are the same as for $u_4$, and so we expect a bound on the
mass of $d_4$ similar to that on $u_4$.  If $m_{d_4} < m_t + m_W$, 
then $d_4$ decay could proceed through a ``doubly-Cabbibo'' suppressed 
tree--level process $d_4 \ra c W$ or through the
one-loop process $d_4 \ra b Z$. 
The relative branching ratios depend on 
details~\cite{Sher:1999ae,Arhrib:2000ct}.
In particular, taking ${\rm BR}(d_4 \ra b Z) = 1$, CDF obtains the bound 
$m_{d_4} > 268$~GeV at 95\% CL~\cite{cdf-bprimeZ}
We choose to adopt the largely CKM-independent bound 
$m_{u_4,d_4} > 258$~GeV throughout.\bigskip

The off-diagonal elements $V_{u_4 i},V_{j d_4}$ of the CKM matrix 
$V = y^u {y^d}^\dagger$ are constrained by flavor physics.
As in the Standard Model, the flavor-violating neutral current 
effects occur in loops and are automatically GIM suppressed.
Rough constraints on the mixing between the first/second 
and fourth generation can be extracted requiring unitarity 
of the enlarged $4\times4$ CKM matrix.  The SM $3 \times 3$ sub-matrix
is well tested by a variety of SM processes~\cite{Yao:2006px}.
The first row of the matrix, combined with measurements of $V_{ud}$,
$V_{us}$, and $V_{cb}$, yields 
\begin{equation}
|V_{u d_4}|^2 = 
1 - |V_{ud}|^2 - |V_{us}|^2 - |V_{ub}|^2 \,\, \simeq \,\, 0.0008 \pm 0.0011 \; .
\end{equation}
For the second row we can use 
the hadronic $W$ branching ratio 
to obtain
\begin{equation}
|V_{c d_4}|^2 = 
1 - |V_{cd}|^2 - |V_{cs}|^2 - |V_{cb}|^2 \,\, \simeq \,\, -0.003 \pm 0.027 \; .
\end{equation}
Similarly, the first column of the matrix allows one to infer,
\begin{equation}
|V_{u_4 d}|^2 = 
1 - |V_{ud}|^2 - |V_{us}|^2 - |V_{ub}|^2 \,\, \simeq \,\, -0.001 \pm 0.005 \; .
\end{equation}
If we require
the above relations be satisfied to $1\sigma$, we obtain
\begin{alignat}{5}
|V_{u d_4}| &\lsim& 0.04 \notag \\
|V_{u_4 d}| &\lsim& 0.08 \notag \\
|V_{c d_4}| &\lsim& 0.17 
\end{alignat}
which are, nevertheless, still significantly larger than the smallest
elements in the CKM matrix $|V_{ub}|$,$|V_{td}|$.
The remainder of the elements ($V_{t d_4}$, $V_{u_4 s}$, $V_{u_4 b}$, 
and $V_{u_4 d_4}$)
could be constrained through a global fit to the
$4 \times 4$ CKM matrix, including the contributions of the 
fourth--generation quarks to specific observables in loops
(for example \cite{Hou:1986ug}),
but this is beyond the scope of this work.  
Similarly there are two additional CP-violating phases in the 
$4\times4$ CKM matrix, but since their effects are proportional to
the unknown (real parts) of the off-diagonal CKM mixings, we ignore their
effects.

The least constrained sector is the mixing between the third and fourth
generations.  The observation of single top 
production~\cite{Abazov:2006gd,cdf-st} can be used to obtain
a lower limit $V_{tb} > 0.68$ at 
$95\%$ C.L.~\cite{Abazov:2006gd}, which still allows for large
third/fourth generation mixing.
Thus it seems likely that fourth generation 
charged-current decays will be mostly into third generation quarks, 
provided the mass difference is large enough to permit 
two-body decays.\bigskip

The new elements in the PMNS matrix $U = y^{\nu} {y^e}^\dagger$ also have 
constraints from lepton flavor violation in the charged and neutral 
sectors.  The most stringent constraint is the absence of  
$\mu \ra e\gamma$.  For weak--scale purely Dirac neutrinos
this constraint~\cite{Cheng:1985bj} implies
$|U_{e 4}U_{\mu 4}| \lsim 4 \times 10^{-4}$.
This suggests that first/second generation mixings with the
fourth generation should be smaller than about $0.02$.
Other generational mixings can also be constrained from 
the absence of lepton flavor violating effects, where again
third/fourth generation mixings are (as expected) the most 
weakly constrained.

There is, however, a significant constraint from
neutrinoless double beta decay on $|U_{i4}|$ in the presence of a  
weak--scale
Majorana mass $M_{44}$.
Such a decay can be mediated by a very light neutrino 
mixing with a weak--scale (partly) Majorana neutrino.
Using
Ref.~\cite{Bamert:1994qh}
and assuming only first/fourth generational mixing,
we obtain
\begin{equation}
\frac{|U_{e4}|^2 p_F^2 M_{44}}{3 m_D^2} \lsim \mathrm{eV} \, ,
\end{equation}
where $m_D = y_{44}^\nu v$ and PMNS phases are ignored.  This expression 
is valid as long as the fourth--generation neutrino masses exceed the 
characteristic energy scale of the double--beta nuclear process, 
$m_{\nu_{1,2}} \gg p_F \simeq 60$~MeV.  Inserting characteristic values,
we obtain
\begin{equation}
|U_{e4}| \lsim 0.9 \times 10^{-2} \; 
\frac{m_D}{M_{44}^{1/2} (100 \; {\rm GeV})^{1/2}}
\end{equation}
No bound remains once the fourth--generation Majorana mass is made 
small, $M_{44} \lsim 10$~MeV. 

\section{Electroweak Constraints}
\label{electroweak-sec}

The most pernicious effect of a fourth generation is the contribution
to oblique electroweak corrections.  $B \leftrightarrow W^3$ mixing is
enhanced, leading to a positive contribution 
$\Delta S = 0.21$ in the limit of degenerate isospin multiplets
(quark and lepton).  Degeneracy is usually assumed for simplicity since split 
doublets significantly contribute to the isospin violating parameter $T$.

There are three important effects that can mitigate the contribution 
to $\Delta S$.  The first, and most important, is exploiting the 
relative experimental insensitivity to the $\Delta S \simeq \Delta T$
direction in oblique parameter space.
We will be more precise below,
but suffice to say slightly split electroweak doublets are 
in far better agreement with electroweak data than without the 
$\Delta T$ contribution.  The second effect involves a reduced
contribution to $S$ by splitting the fourth--generation multiplets
in a particular mass hierarchy.  The last, and least important effect 
is introducing a Majorana mass for the fourth--generation neutrino.\bigskip

Splitting the up-type from down-type fermion masses in the same electroweak
doublet can give a negative contribution to $S$.  In the large mass limit 
$m_{u,d} \gg M_Z$, the contribution to $S$ depends logarithmically 
on the ratio $m_u/m_d$~\cite{Peskin:1991sw,He:2001tp}
\begin{equation}
\Delta S = \frac{N_c}{6 \pi} \left( 1 - 2 Y \ln \frac{m_u^2}{m_d^2} \right)
\label{S-eq}
\end{equation}
where $Y$ is the hypercharge of the left-handed doublet of fermions
with degeneracy (color factor) $N_c$.
Clearly the fourth--generation contributions to $S$ are 
reduced if $m_{u_4}/m_{d_4} > 1$ for quarks ($Y = 1/6$)
and $m_{\nu}/m_{\ell} < 1$ for leptons $(Y = -1/2)$.
How big can this effect be given that split multiplets also
contribute to $\Delta T$?

%%%%%%%%%%%%%%%%%%%%%%%%%%%%%%
\begin{figure}[t]
\includegraphics[width=1.0\hsize]{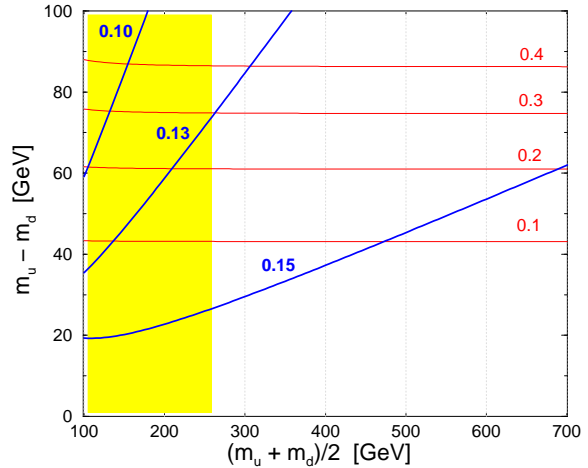}
\vspace*{-10mm}
\caption{Contours of constant $\Delta S_q$ (diagonal, blue) 
and $\Delta T_q$ (horizontal, red) for the fourth--generation quarks.  
The plot is symmetric with respect
to $m_{u_4} - m_{d_4} \leftrightarrow m_{d_4} - m_{u_4}$, since $\Delta T_q$ 
is positive definite.  The Tevatron bound 
$m_{u_4,d_4} > 258$~GeV excludes the shaded (yellow) region.}
\label{quarksST-fig}
\end{figure}
%%%%%%%%%%%%%%%%%%%%%%%%%%%%%%

To calculate $\Delta S$ (and $\Delta T$ and $\Delta U$) we use
exact one-loop expressions
which are valid for all $m_{u,d}$~\cite{Kniehl:1990mq}. 
We checked our formulae by   
explicitly verifying finiteness (renormalization scale independence)
as well as finding numerical agreement with several explicit results
given in Ref.~\cite{He:2001tp}.
In Fig.~\ref{quarksST-fig}
we show the size of the contribution from the $(u_4,d_4)$ doublet
as a function of the masses of the quarks.  
The effect of using the exact one--loop expressions is modest; 
in fact Eq.~(\ref{S-eq}) reproduces the $S$ contours 
up to an accuracy $\pm 0.01$ throughout the plot.
The typical size of $U$ is smaller than $0.02$ 
everywhere, and so we set $U=0$ throughout.

For the leptons, what is most important is the split 
between the neutral and charged fermion masses.
For example,
$m_{\nu, \ell} \simeq 100, 135$~GeV implies
$(\Delta S_\nu,\Delta T_\nu) \simeq (0.02,0.02)$, and the slightly larger
values 
$m_{\nu, \ell} \simeq 100, 155$~GeV give
$(\Delta S_\nu,\Delta T_\nu) \simeq (0.00,0.05)$.
These results from the exact one-loop formulae agree surprisingly
well with Eq.~(\ref{S-eq}), despite the lepton masses being 
near $M_Z$.\bigskip

Fits of the combined electroweak data provide 
constraints on the oblique parameters and have been performed by 
the LEP Electroweak Working Group (LEP~EWWG)~\cite{Alcaraz:2006mx}
and separately by the PDG~\cite{Yao:2006px}.  Both fits
find that the Standard Model defined by $(S,T)=(0,0)$ with 
$m_t = 170.9$~GeV and $m_H = 115$~GeV is within $1\sigma$
of the central value (always holding $U=0$).  
However, the two fits disagree on the best--fit point.  
The latest LEP~EWWG fit finds a central value 
$(S,T)=(0.06,0.11)$~\cite{howwegotthis} with a $68\%$ contour 
that is elongated along the $S \simeq T$ major axis from 
$(S,T)=(-0.09,-0.03)$ to $(0.21,0.25)$. 
By contrast, the PDG find the central value $(S,T)=(-0.07,-0.02)$ 
after adjusting $T$ up by $+0.01$ to account for the latest value 
of $m_t = 170.9$~GeV.

The most precise constraints on $S$ and $T$ arise from 
$\sin^2 \theta_{\rm lept}^{\rm eff}$ and $M_W$, used by both groups.
The actual numerical constraints derived from these measurements
differ slightly between each group,
presumably due to slight updates of data (the $S$-$T$ plot generated by the 
2006 LEP~EWWG is one year newer than the plot 
included in the 2006 PDG).  A larger difference concerns the use of the 
$Z$ partial widths and $\sigma_h$.  The LEP~EWWG
advocate using just $\Gamma_{\ell}$, since it is insensitive to
$\alpha_s$. This leads to a flatter constraint in the $S$-$T$ plane.  
The PDG include the $\alpha_s$-sensitive quantities $\Gamma_Z$, $\sigma_h$, 
$R_q$ as well as $R_{\ell}$, and obtain a less flat, more oval-shaped 
constraint.
Additional lower--energy data can also be used to (much more weakly) 
constrain $S$ and $T$, although there are systematic uncertainties
(and some persistent discrepancies 
in the measurements themselves).  
The LEP~EWWG do not include lower--energy data in their fit, 
whereas the PDG appear to include some of it.
In light of these subtleties, we choose to use the 
LEP~EWWG results when quoting
levels of confidence of our calculated shifts in the $S$-$T$ plane.
We remind the reader, however, that the actual level of confidence
is obviously a sensitive function of the precise nature of the
fit to electroweak data.\bigskip

%%%%%%%%%%%%%%%%%%%%%%%%%%%%%%
\begin{table}[t]
\begin{tabular}{c|ccc|cc}
parameter set & $m_{u_4}$ & $m_{d_4}$ & $m_H$ & $\Delta S_{\rm tot}$ 
  & $\Delta T_{\rm tot}$ \\ \hline
(a) & $310$ & $260$ & $115$ & $0.15$ & $0.19$ \\
(b) & $320$ & $260$ & $200$ & $0.19$ & $0.20$ \\
(c) & $330$ & $260$ & $300$ & $0.21$ & $0.22$ \\ \hline
(d) & $400$ & $350$ & $115$ & $0.15$ & $0.19$ \\ 
(e) & $400$ & $340$ & $200$ & $0.19$ & $0.20$ \\ 
(f) & $400$ & $325$ & $300$ & $0.21$ & $0.25$ \\ 
\end{tabular}
\caption{Examples of the total contributions to $\Delta S$ and 
$\Delta T$ from a fourth generation.
The lepton masses are fixed to $m_{\nu_4} = 100$~GeV and 
$m_{\ell_4} = 155$~GeV, giving $\Delta S_{\nu \ell} = 0.00$
and $\Delta T_{\nu \ell} = 0.05$.  The best fit to data is 
$(S,T)=(0.06,0.11)$~\cite{howwegotthis}.  
The Standard Model is normalized to $(0,0)$ for 
$m_t = 170.9$~GeV and $m_H = 115$~GeV.  All points are
within the 68\% CL contour defined by the 
LEP~EWWG~\cite{howwegotthis}.}
\label{tab:params}
\end{table}
%%%%%%%%%%%%%%%%%%%%%%%%%%%%%%

\begin{figure}
\begin{center}
\includegraphics[width=1.0\hsize]{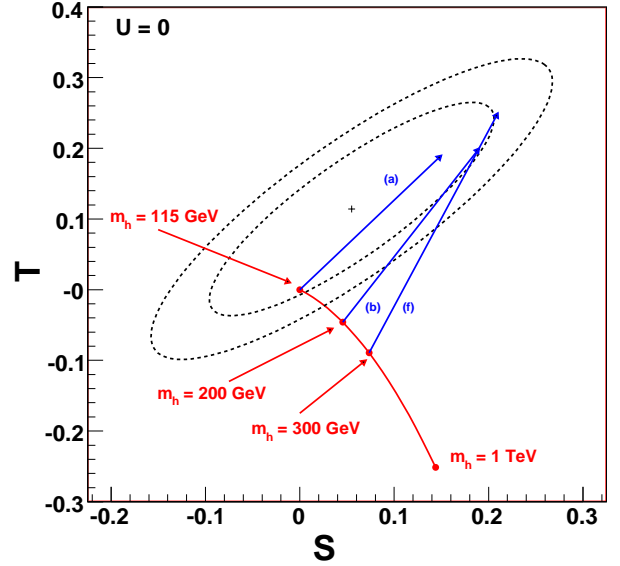}
\end{center}
\caption{The 68\% and 95\% CL constraints on the $(S,T)$ parameters
obtained by the LEP Electroweak Working Group
\cite{Alcaraz:2006mx,howwegotthis}.  The shift in $(S,T)$ resulting
from increasing the Higgs mass is shown in red.  The shifts in
$\Delta S$ and $\Delta T$ from a fourth generation with several of
the parameter sets given in Table \ref{tab:params} are shown in blue.}
\label{ST-fig}
\end{figure}

In Table~\ref{tab:params} we provide several examples of
fourth--generation fermion masses which yield contributions
to the oblique parameters that are within the $68\%$ CL ellipse
of the electroweak precision constraints.
We illustrate the effect of increasing Higgs mass with
compensating contributions from a fourth generation in
Fig.~\ref{ST-fig}.
%%%%%%%%%%%%%%%%%%%%%%%%%%%%%%
%%%%%%%%%%%%%%%%%%%%%%%%%%%%%%
More precisely, the fit to electroweak data is in agreement with
the existence of a fourth generation and a light Higgs about as well as
the fit to the Standard Model alone with $m_H = 115$~GeV.
Using suitable contributions from the fourth--generation quarks, 
heavier Higgs masses up to $315$~GeV remain in agreement with the 
$68\%$ CL limits derived from electroweak data.  Heavier Higgs 
masses up to 750~GeV are permitted if the agreement with data 
is relaxed to the $95\%$ CL limits.\bigskip

Until now we have focused on purely Dirac neutrinos. However,
there is also a possible reduction of $S_{\rm tot}$ when the 
fourth--generation neutrino has a Majorana mass comparable to the
Dirac mass~\cite{Gates:1991uu,Kniehl:1992ez}.
Using the exact one-loop expressions of Ref.~\cite{Kniehl:1992ez},
we calculated the contribution to the electroweak parameters with 
a Majorana mass.
Given the current direct--search bounds from LEP~II
on unstable neutral and charged leptons, we find a Majorana mass 
is unfortunately not particularly helpful in significantly lowering $S$.  
A Majorana mass does, however, enlarge the parameter space where
$S \simeq 0$.
For example, given the lepton Dirac and Majorana masses
$(m_D,M_{44}) = (141,100)$~GeV, the lepton mass eigenstates are
$(m_{\nu_1},m_{\nu_2},m_{\ell}) = (100,200,200)$~GeV, 
and contributions to the oblique parameters of
$(\Delta S_\nu,\Delta T_\nu) = (0.01,0.04)$.
It is difficult to find parameter regions with $\Delta S_{\ell} < 0$ 
without either contributing to $\Delta U_{\ell} \simeq - \Delta S_{\ell}$, 
contributing significantly more to $\Delta T_{\ell}$, 
or taking $m_{\nu_1} < 100$~GeV which violates the LEP~II bound
for unstable neutrinos.\bigskip

Let us summarize our results thus far. We have identified a region
of fourth--generation parameter space in agreement with all 
experimental constraints and with 
minimal contributions to the electroweak precision
oblique parameters.  This parameter space is characterized by
\begin{alignat}{5}
m_{\ell_4} - m_{\nu_4} &\simeq  30 - 60 \; \mathrm{GeV} \notag \\
m_{u_4} - m_{d_4}  &\simeq  
  \left( 1 + \frac{1}{5} \ln \frac{m_H}{115 \, \mathrm{GeV}} 
                         \right) \times 50 \; \mathrm{GeV} \notag \\
|V_{u d_4}|,|V_{u_4 d}| &\lsim 0.04 \notag \\
|U_{e 4}|,|U_{\mu 4}|   &\lsim 0.02 \; ,
\label{eq:benchmarks}
\end{alignat}
and subject to the current direct search limits
$m_{\nu_4,\ell_4} > 100$ GeV and
$m_{u_4,d_4} > 258$ GeV.
The other elements of the CKM and PMNS matrix are not
strongly constrained.
The smallest contribution to the oblique parameters 
occurs for small Higgs masses. 
The leptons and quark masses are not significantly split,
in particular, the two--body decays $\ell_4 \ra \nu_4 W$ and
$d_4 \ra u_4 W$ generally do not occur.
Finally, while there are strong restrictions on the mass
\emph{differences} between the up-type and down-type fields,
there are much milder restrictions on the \emph{scale} of the
mass.  

\section{Higgs Searches}
\label{higgs-sec}

The set of mixing elements and mass hierarchies shown in 
Eq.(\ref{eq:benchmarks}) has significant effects on Higgs searches 
at the Tevatron and at the LHC.  One clear observation is that
Higgs decays into fourth--generation particles, if possible at all, 
are expected only into leptons, unless the Higgs is exceptionally heavy
which is disfavored by precision data.
%Given the mass hierarchy, we now make a series of sharp predictions 
%for the effects of a fourth generation on Higgs physics.

A fourth generation with two additional heavy quarks is well known 
to increase the 
effective $ggH$ coupling by roughly a factor of $3$, and hence 
to increase the production cross section $\sigma_{gg \to H}$ by a factor 
of roughly 9~\cite{Gunion:1994zm}.  
The Yukawa coupling 
exactly compensates for the large decoupling quark masses in the 
denominator of the loop integral~\cite{Ellis:1975ap}.
This result is nearly independent of the mass of the heavy quarks,
once they are heavier than the top. 
(Modifications to the Higgs production cross section has also been 
considered in an effective theory approach in Ref.~\cite{Manohar:2006gz}.)
This enhancement allowed CDF and D0 to very recently rule out a Higgs
in a four generation model within the mass window of roughly 
$145 < m_H < 185$~GeV to 95\% CL using the process 
$gg \ra h \ra W^+W^-$~\cite{Hsu:2007qq,d0-higgs-fourth}. 
While over recent years weak--boson--production has
proven the leading discovery channels for light Higgs bosons
--- in the Standard Model as well as in extensions with more than
one Higgs doublet, like for example the MSSM~\cite{Plehn:1999nw} ---
these channels are less promising in models with a fourth generation, because
the loop effects on the $WWH$ couplings are small enough to be ignored 
in the Standard Model.\bigskip

The increase in the $ggH$ coupling dramatically increases the 
decay rate of $H \ra gg$.  For Higgs masses lighter than about 
$140$~GeV and no new two--body decays, this decay dominates, 
but is probably impossible to extract from the two-jet background 
at the LHC.
The presence of this decay effectively suppresses all other two--body decays, 
including the light--Higgs discovery mode $H \to \tau\tau$, by
roughly a factor $0.6$.  Only once the tree-level decay mode 
$H \to WW^*$ opens does this suppression vanish. 
More subtle effects occur for the loop--induced decay
$H \ra \gamma\gamma$. 
The partial widths for $H \to \gamma \gamma$ and $H \to gg$ can be 
written as~\cite{Ellis:1975ap}
\begin{alignat}{5}
 \Gamma_{H \to \gamma \gamma} &=  \frac{G_\mu \alpha^2 m_H^3}{128
  \sqrt{2} \pi^3} \left| \sum_{f} N_c Q_f^2 A_f(\tau_f) + A_W(\tau_W)
  \right|^2
                \notag \\
  \Gamma_{H \to g g} &=  \frac{G_\mu \alpha_s^2 m_H^3}{36
  \sqrt{2} \pi^3} \left| \frac{3}{4} \sum_{f}  A_f(\tau_f) \right|^2 \; .
\label{eq:hgg}
\end{alignat}
where $A_f$ and $A_W$ are the form factors for the spin-$\frac{1}{2}$ and
spin-$1$ particles respectively.  These form factors are
\begin{alignat}{5}
 A_f(\tau) &= 2 \left[  \tau 
                      + (\tau - 1) f(\tau) 
                \right] \, \tau^{-2} \notag \\
 A_W(\tau) &= - \left[  2 \tau^2 
                      + 3 \tau + 3(2 \tau -1) f(\tau)
                \right] \, \tau^{-2}
\end{alignat}
with $\tau_i = m_H^2/4 m_i^2$, $(i=f,W)$ and $f(\tau)$ defined as
the three--point integral
\begin{alignat}{5} 
f(\tau)= \left\{
                 \begin{array}{cc}
                 \mathrm{arcsin}^2 \sqrt{\tau} & \tau \le 1 \\
                -\frac{\displaystyle 1}{\displaystyle 4} 
                             \left[ \ln \frac{\displaystyle 
                                               1+\sqrt{1-\tau^{-1}}}
                                              {\displaystyle 
                                               1-\sqrt{1-\tau^{-1}}}
                                    -i \pi 
                             \right]^2        & \tau > 1
\end{array}
\right.
\end{alignat}
%

%%%%%%%%%%%%%%%%%%%%%%%%%%%%%%
\begin{table}[t]
\begin{tabular}{c|c|c}
$m_H$         & $115$              & $200$                 \\  \hline
$A_W$         & $-8.0321$          &   $-9.187-5.646 i$    \\
$A_t$         & $1.370$            &   $1.458$             \\
$A_{u_4}$     & $1.344$            &   $1.367$             \\
$A_{d_4}$     & $1.349$            &   $1.382$             \\
$A_{\ell_4}$  & $1.379$            &   $1.491$             \\
\end{tabular}
\caption{The dominant form factors for the decay 
$H \to \gamma \gamma$ and $H \to gg$ according to 
Eq.(\ref{eq:hgg}) for the parameter points (a) and (b). 
For $H \to gg$ just the quark loops contribute.
The form factors are obtained from a modified version of 
Hdecay~\cite{Djouadi:1997yw}.}
\label{tab:formfactors}
\end{table}
%%%%%%%%%%%%%%%%%%%%%%%%%%%%%%

From the numbers given in Table~\ref{tab:formfactors} we see that the
$ggH$ coupling indeed consists of nearly identical contributions from
the SM top quark and the two additional fourth--generation quarks. In
particular, the contributions of the fourth--generation quarks in the
parameters points (a) and (b) are well described by the
decoupling limit in which we estimated the enhancement of the Higgs
production rate as a factor of 9. For a 200~GeV Higgs we start to
observe very small top--mass effects.  This means that the enhancement
factor in $\sigma_{gg}$ slowly decreases from $8.5$ to $7.7$ for Higgs
masses between 200 and 300~GeV.  Of course, this scaling factor breaks
down for the top threshold region around 350~GeV and subsequent
heavy-quark thresholds. This corresponds to the absorptive imaginary
parts of the $A_i$ listed in Table~\ref{tab:formfactors}.

In the Standard Model the Higgs decay to photons is dominated by the
$W$ loop, which destructively interferes with the smaller top--loop.
In Table~\ref{tab:formfactors} we see how in the fourth--generation
model all additional heavy particles contribute to the loop. For a
light Higgs boson this implies a suppression of the branching ratio
${\rm BR}(\gamma\gamma)$ by roughly a factor
1/9~\cite{footnote4}.  Suppression of the $h \ra \gamma\gamma$ mode
has also been recently considered in a somewhat different context
in Ref.~\cite{Phalen:2006ga}.\bigskip

%%%%%%%%%%%%%%%%%%%%%%%%%%%%%%
\begin{figure}[t]
\includegraphics[width=1.0\hsize]{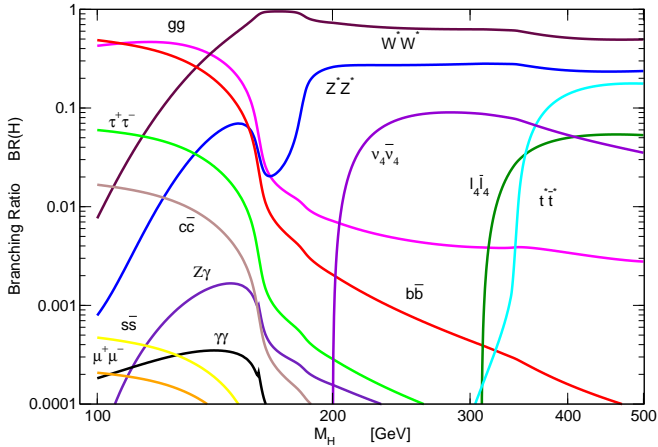}
\vspace*{-5mm}
\caption{Branching ratio of the Higgs with fourth--generation effects
assuming $m_{\nu} = 100$~GeV and $m_{\ell} = 155$~GeV.  The loop effects to 
$H \ra gg$ and $H \ra \gamma\gamma$ are largely insensitive to the 
fourth--generation quark masses. For the fourth--generation masses we follow
the reference point (b).}
\label{BR-fig}
\end{figure}
%%%%%%%%%%%%%%%%%%%%%%%%%%%%%%

We show the complete set of branching ratios in Fig.~\ref{BR-fig}.
All predictions for Higgs decays are computed with a modified version
of Hdecay~\cite{Djouadi:1997yw} which includes radiative corrections
also to the fourth--generation decays, but no off-shell effects for
these decays. The two thresholds in ${\rm BR}(\ell_4 \bar{\ell}_4)$
and ${\rm BR}(\nu_4 \bar{\nu}_4)$ compete with the larger top decay
channel with its color factor $N_c$, but all of them are small
compared to the gauge boson decays.  Higgs decays to
fourth--generation quarks are implemented in the extended version of
Hdecay but only occur for larger Higgs masses.

For a light Higgs below 200~GeV the effects on different gluon--fusion
channels are roughly summarized by
\begin{alignat}{5}
     \sigma_{gg} {\rm BR}(\gamma\gamma) \Big|_{\rm G4} \; \simeq \;
   & \sigma_{gg} {\rm BR}(\gamma\gamma) \Big|_{\rm SM} \notag \\
     \sigma_{gg} {\rm BR}(ZZ) \Big|_{\rm G4} \; \simeq \;
   & (5 \cdots 8) \;
     \sigma_{gg} {\rm BR}(ZZ) \Big|_{\rm SM} \notag \\
     \sigma_{gg} {\rm BR}(f\overline{f}) \Big|_{\rm G4} \; \simeq \;
   & 5 \; 
     \sigma_{gg} {\rm BR}(f\overline{f}) \Big|_{\rm SM} 
\end{alignat}

\begin{figure}[t]
\includegraphics[width=1.0\hsize]{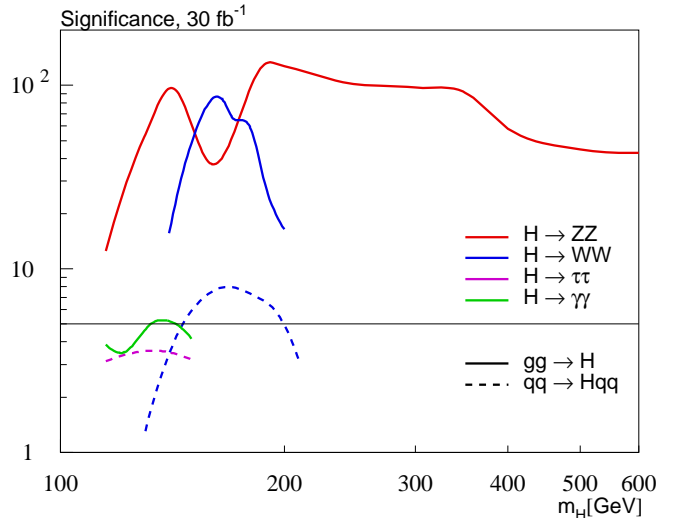}
\vspace*{-5mm}
\caption{Scaled LHC discovery contours for the fourth--generation
 model. All channels studies by CMS are included. The significances 
 have naively been scaled to the modified production rates and 
 branching rations using the fourth--generation parameters 
 of reference point (b).}
\label{signif-fig}
\end{figure}

In Figure~\ref{signif-fig} we show a set of naively scaled discovery
contours for a generic compact LHC detector, modifying all known
discovery channels according to fourth--generation effects~\cite{cms}.
The enhancement of the production cross section implies the the
``golden mode'' $H \ra ZZ \to 4\mu$ can be used throughout the Higgs
mass range, from the LEP~II bound to beyond 500~GeV.  Both $WW$
channels~\cite{Rainwater:1999sd,Dittmar:1996ss} are still relevant,
but again the gluon--fusion channel (which in CMS analyses for a SM
Higgs tends to be more promising that the weak--boson--channel, while
Atlas simulation show the opposite~\cite{Buscher:2005re}) wins due to
the fourth--generation enhancement.  As mentioned above, the
weak--boson--fusion discovery decay $H \to \tau \bar{\tau}$ becomes
relatively less important, even though its significance is only
slightly suppressed.  Weak--boson--fusion production with a subsequent
decay to photons is suppressed by one order of magnitude compared to
the Standard Model and not shown anymore, while for the gluon--fusion
channel with a decay to photons the corrections to the production rate
and the decay width accidentally cancel.

%%%%%%%%%%%%%%%%%%%%%%%%%%%%%%
%%%%%%%%%%%%%%%%%%%%%%%%%%%%%%

Measuring the relative sizes of the different production and decay
modes would allow an interesting study of Higgs properties that should
be easily distinguishable from other scenarios (two Higgs doublet
model, supersymmetry, etc.).  Moreover, there may be novel search
strategies for the Tevatron that would be otherwise impossible given
just the SM Higgs production rate.\bigskip

%%%%%%%%%%%%%%%%%%%%%%%%%%%%%%
\begin{figure}[t]
\includegraphics[width=1.0\hsize]{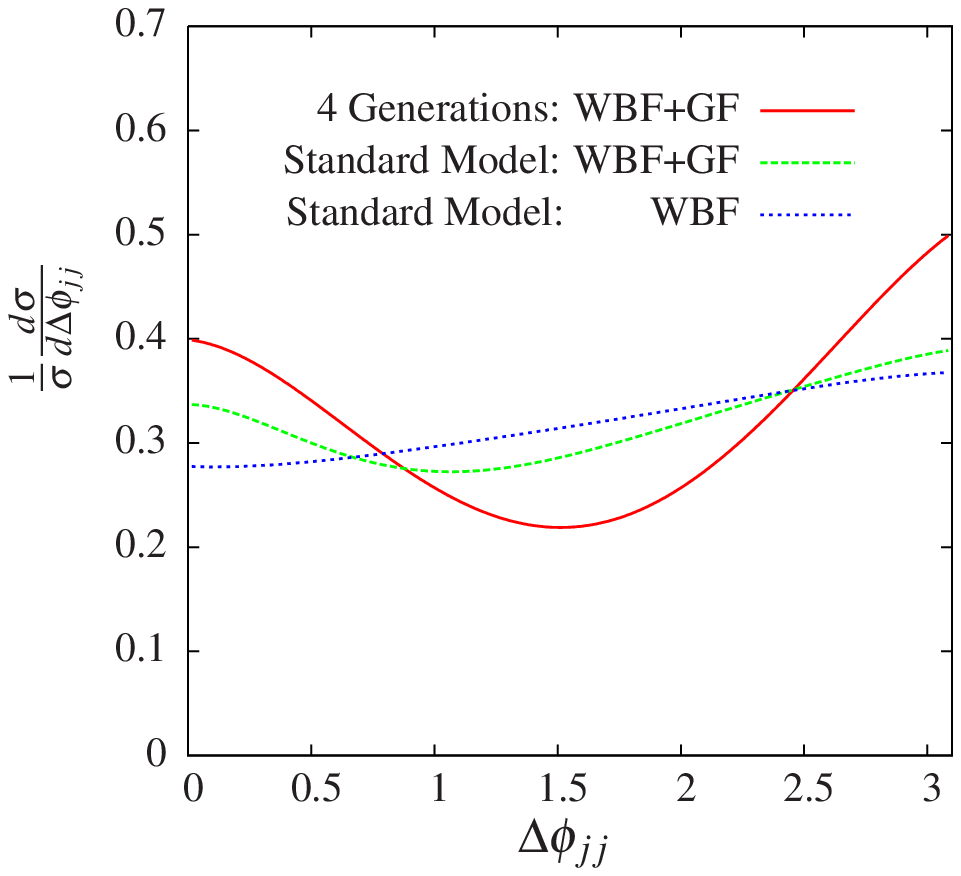}
\vspace*{-5mm}
\caption{Angular distribution of vector-boson fusion channel
 assuming reference point (b) with its Higgs mass $m_H=200$~GeV }
\label{angular-fig}
\end{figure}
%%%%%%%%%%%%%%%%%%%%%%%%%%%%%%

Weak--boson--fusion Higgs production has interesting features beyond
its total rate. Most importantly, it has the advantage of allowing us
to extract a Higgs sample only based on cuts on the two forward
tagging jets, allowing us to observe Higgs decays to taus and even
invisible Higgs decays~\cite{Plehn:1999nw,Eboli:2000ze}. Among the
relevant distribution for this strategy are the angular correlations
between the tagging jets: for two $W$ bosons coupling to the Higgs
proportional to the metric tensor we find that the azimuthal angle
correlation between the tagging jets is flat, modulo slight effects of
the acceptance cuts. For a coupling to the Higgs proportional to the
transverse tensor the same distribution peaks around $\Delta \phi_{jj}
= 0,\pi$. This correlation can be used to determine the Lorentz
structure of the $WWH$ coupling~\cite{Plehn:2001nj}.

The modification to the $ggH$ coupling from a fourth generation leads
to a larger relative size of the gluon--fusion process in the
$H$+2~jets sample.  This causes a modification in the angular
correlation. shown in Fig.~\ref{angular-fig}. For our
Madevent~\cite{Alwall:2007st} simulation we employ the cuts listed in
Ref.~\cite{DelDuca:2001ad} with $m_{jj} > 600$~GeV and use the HEFT
model~\cite{Ellis:1975ap}.  Measuring this distribution would 
provide an interesting probe
of the relative sizes of the weak vector boson fusion over gluon
fusion. Of course this relative weight will be affected by cuts as
well as analysis strategies like a mini--jet veto and requires a
careful study.\bigskip

New decay modes of the Higgs are possible if the Higgs is sufficiently
heavy.  Simply trying to produce the Higgs and decay to two heavy quarks
at hadron colliders is small compared with the QCD production and therefore
not promising.  For decays to heavy leptons there are two cases to
distinguish, depending on the size of the mixing between the
fourth--generation leptons and the SM leptons.

One very interesting modification to Higgs signals occurs if 
the mixing between the fourth--generation leptons and the other 
generations is very small ($|U_{i4}| < 10^{-8}$).
In this case, the fourth--generation neutrinos escape the detector 
as missing energy.  This will be the case, for example, when one
contemplates the fourth--generation neutrino as dark matter.
(The intermediate case of decay with a displaced vertex is also possible 
for a narrower range of PMNS mixings of roughly
$10^{-6} \lsim |U_{i4}| \lsim 10^{-8}$.
A recent discussion of the possibility of displaced vertices 
associated with Higgs decay to neutrinos, in a different context, 
can be found in \cite{Graesser:2007yj}.)
LEP~II bounds on
missing energy plus an initial--state photon are relatively weak,
and thus the fourth--generation neutrino can be as light as about $M_Z/2$.  
This case also requires a mechanism to avoid the direct detection 
bounds (we comment on this below) which otherwise rule out 
weak scale Dirac neutrinos as dark matter. 
For Higgs masses below 140~GeV, 
the invisible decay $H \ra \nu_4\overline{\nu}_4$ can even dominate.
Such a signature is among the more challenging at the LHC, 
in particular because the most likely 
channel to observe an invisible Higgs is weak boson fusion, which is not
enhanced by fourth--generation loop effects~\cite{Eboli:2000ze}.

If the mixing $|U_{i4}|$ is not exceedingly small, then the 
fourth--generation neutrino promptly decays via an PMNS mixed charged 
current $U_{i4} \ell_i^\pm \nu_4 W^\mp$.  Given the LEP bounds 
for this two--body decay to be open,
the Higgs must be heavier than about 200~GeV.  This means that
the new signal is $H \ra \nu_4\overline{\nu}_4 \ra \ell^+\ell^-W^+W^-$
where the lepton flavor depends on which PMNS mixing element dominates.
The branching ratio of this mode, shown in Fig.~\ref{BR-fig}, is roughly
5\% for Higgs masses larger than the kinematic threshold.
When combined with the branching ratio of the $W$'s into leptons,
we can estimate that the rate into four leptons (plus missing energy)
\begin{equation}
\frac{{\rm BR}(H \ra \nu_4\overline{\nu}_4 \ra 4\ell)}{
{\rm BR}(H \ra ZZ \ra 4\ell)} \simeq 1.1 \left(
\frac{{\rm BR}(H \ra \nu_4\overline{\nu}_4)}{0.1} \right)
\end{equation}
Hence, the rate is comparable to the rate for $H \ra ZZ \ra 4\ell$.
One subtlety is that the decay $\nu_4 \ra \ell W$ likely proceeds
to third generation leptons, if indeed the PMNS mixing element 
$|U_{\tau4}|$ is largest, and so the
two leptons from this decay would be $\tau$'s.
It might nevertheless be worthwhile to study the four lepton signal
characteristics, including the relative rates into different lepton flavors,
as well as searching for events with accompanying missing energy.

In the case where the fourth--generation neutrino has an
electroweak scale Majorana mass, $M_{44} \sim v y^{\nu}_{44}$,
half of the time the same two--body decay proceeds to same-sign leptons
$H \ra \nu_4\nu_4 \ra \ell^\pm \ell^\pm W^\mp W^\mp$.
This is a rather unusual signal of the Higgs has little physics background, 
except potentially Higgs pair production, with each Higgs decaying into 
$W$ pairs. The difference is that the four generation signal has 
no missing energy, and moreover, the visible mass of the events
would approximately reconstruct the Higgs mass and
not threshold--suppressed two--Higgs production.\bigskip

%%%%%%%%%%%%%%%%%%%%%%%%%%%%%%
\begin{table}[t]
\begin{tabular}{cc|r|rr}
  & $\lambda_{HHH}$ & $m_H$ & $\sigma_{gg \to HH}$ 
                    & $\sigma_{gg \to HH} {\rm BR}(4W)$ \\ \hline
  SM  & $\lambda_{\rm SM}$ & 115 & 34.07 &   0.22 \\
  SM  & 0                  & 115 & 63.56 &   0.41 \\
  SM  & $\lambda_{\rm SM}$ & 200 &  8.54 &   4.61 \\
  SM  & 0                  & 200 & 25.73 &  13.89 \\
  (a) & $\lambda_{\rm SM}$ & 115 & 299.7 &   0.76 \\
  (a) & 0                  & 115 & 500.2 &   1.26 \\
  (b) & $\lambda_{\rm SM}$ & 200 &  96.2 &  51.30 \\
  (b) & 0                  & 200 & 241.3 & 128.6
\end{tabular}
\caption{Total cross section for Higgs pair production at the LHC
 for two different
 Higgs masses, 115~GeV and 200~GeV according to reference points (a) and (b).
 All masses are given in units of GeV, all rates in units of fb.
}
\label{tab:pairs}
\end{table}
%%%%%%%%%%%%%%%%%%%%%%%%%%%%%%

Finally, Higgs pair production is resurrected by fourth--generation
loop effects.  While the SM production rate at the LHC might barely be
sufficient to confirm the existence of a triple Higgs coupling
$\lambda_{HHH}$ as predicted by the Higgs
potential~\cite{Baur:2002rb}, the enhancement of the effective $ggH$
and $ggHH$ couplings should allow for a proper measurement of
$\lambda_{HHH}$.  Enhancements to Higgs pair production using an
operator approach was also recently considered in Ref.~\cite{Pierce:2006dh}.

Total rates are notoriously difficult observables at hadron colliders,
but the Higgs self coupling can be beautifully extracted from the
threshold behavior of the $gg \to HH$ amplitude. At threshold, this
process is dominated by the two form factors $F_{\Delta, \Box}$
proportional to the metric tensor, which arise from the triangular and
box diagrams (following the notation of Ref.~\cite{Plehn:1996wb}).
In the low--energy limit~\cite{Ellis:1975ap} the box diagram's form
factor proportional to the transverse tensor is suppressed by powers
of the loop mass. The Higgs--coupling analysis makes use of the fact
that at threshold the two contributions $F_\Delta$ and $F_\Box$
cancel. More precisely, in the low--energy limit $m_H \ll \sqrt{s} \ll m_t$ 
we find $F_\Delta = - F_\Box + \mathcal{O}(\hat{s}/m_t^2)$. 
This cancellation explains the increase in rate when we set
$\lambda_{HHH}$ to zero, as shown in Table~\ref{tab:pairs}.

\begin{figure}[t]
\includegraphics[width=1.0\hsize]{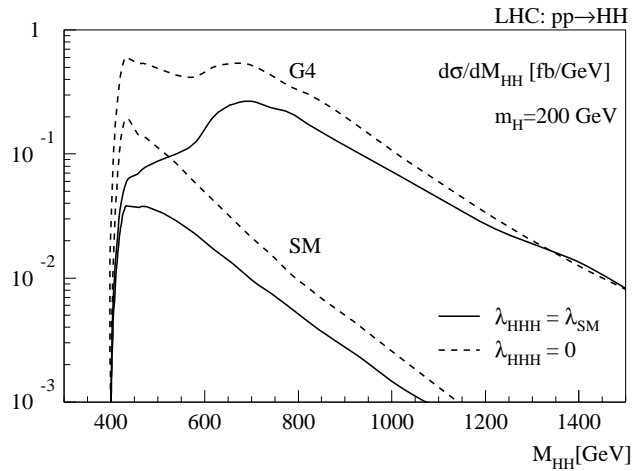}
\vspace*{-5mm}
\caption{Invariant mass distribution for Higgs pair production at the
  LHC. We show the Standard Model and fourth--generation curves in the
  reference point (b). For the dashed line the Higgs self coupling is
  set to zero.}
\label{dihiggs-fig}
\end{figure}

If we only slightly vary the size of the Higgs self coupling, this
threshold behavior changes significantly~\cite{Baur:2002rb} and
provides an experimental handle on $\lambda_{HHH}$. In
Figure~\ref{dihiggs-fig} we show the $HH$ invariant mass (or $\hat{s}$
at parton level) distribution. The shift between finite and zero
$\lambda_{HHH}$ in the Standard Model provides the (S)LHC measurement
of the Higgs self coupling. Similarly to the $ggH$ form factors shown
in Table~\ref{tab:formfactors} the decoupling assumption for top
quarks is numerically not quite as good as for the additional
fourth--generation quarks. Once the process is dominated by heavier
quarks the variation of $m_{HH}$ with $\lambda_{HHH}$ becomes
significantly more pronounced, so there is little doubt that we can
use it to measure the Higgs self coupling.

For the Standard Model, the Higgs self
coupling analysis at the LHC is likely restricted to the $4W$ decay
channel~\cite{Baur:2002rb}. From Table~\ref{tab:pairs} we see that for
light Higgs masses this decay is strongly suppressed, so it would be
an interesting exercise to see if there are alternative decay
channels~\cite{Baur:2003gp} which might work for lighter Higgs bosons,
given the rate and $m_{HH}$ sensitivity increase by the fourth
generation.

%%%%%%%%%%%%%%%%%%%%%%%%%%%%%%
%%%%%%%%%%%%%%%%%%%%%%%%%%%%%%

\section{Meta-stability and Triviality}
\label{stabtriv-sec}

Until now we have concentrated on collider effects of a 
fourth generation coupled to one Higgs doublet.  Since the 
Yukawa couplings of the new fermions exceed $1.5$ 
for the fourth--generation quarks, the four--generation model as an 
effective theory breaks down at a scale that may not be far above
the TeV scale.  There are two well-known constraints:  (1) the possibility
that the quartic coupling is driven negative, destabilizing
the electroweak scale by producing large field minima through quantum
corrections~\cite{Sher:1988mj}, and (2) large Yukawa couplings 
driving the Higgs quartic and/or the Yukawas themselves
to a Landau pole, {\it i.e.} entering a strong--coupling regime.

In both cases the problematic coupling is the Higgs quartic, 
since it receives much larger new contributions to its 
renormalization group running from the fourth--generation 
quark Yukawas couplings.
The renormalization group equation for $\lambda(\mu)$ is
\begin{equation}
16 \pi^2 \frac{d\lambda}{dt} = 12 \lambda^2 - 9 \lambda g_2^2
- 3 \lambda g_1^2 + 4 \lambda \sum N_f y_f^2 -4  \sum N_f y_f^4
\end{equation}
where we have shown only the dominant terms.  The last two terms
encode the Higgs wave function and quartic terms induced by the 
fermions; the sum is over all identical fermions with degeneracy $N_f$.
In our numerical estimations we also include the sub-leading 
electroweak coupling dependence, and evolve using
the full set of one loop $\beta$-functions~\cite{Arason:1991ic}.

We can estimate the scale at which the meta-stability bound becomes
problematic by requiring that the probability of tunneling into 
another vacuum over the current age of the Universe 
is much less than 1. This is equivalent
to the requirement that the running quartic interaction 
is~\cite{Isidori:2001bm}
\begin{equation}
\lambda(\mu) \lsim \frac{4 \pi^2}{3 \ln \left( H/\mu \right)} \; ,
\end{equation}
where $H$ is the Hubble scale.  The scale at which this inequality is
saturated is a minimum scale where new physics is required.  
We should emphasize that the new physics does \emph{not} need to be 
strongly coupled.  For example, a supersymmetric model with four generations
does not have a running quartic that turns negative 
as long as superpartners are (roughly) below the TeV scale.
This is important because weakly coupled physics with particles
obtaining their mass through {\it e.g.} supersymmetry breaking, not 
electroweak breaking, will hardly affect our Higgs results.\bigskip

%%%%%%%%%%%%%%%%%%%%%%%%%%%%%%
\begin{figure}[t]
\includegraphics[width=1.0\hsize]{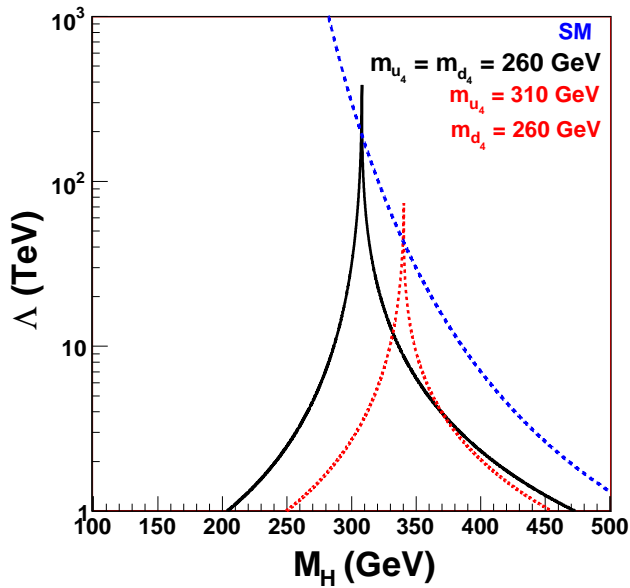}
\vspace*{-10mm}
\caption{The minimum scale at which new physics enters into the 
Higgs potential to avoid either a too short--lived vacuum 
or to avoid a Landau pole in $\lambda$.
These two constraints are qualitatively distinct: meta--stability
can be restored by \emph{weakly} coupled physics below
a TeV scale, whereas the Landau pole signals a strongly interacting
Higgs sector.  The dashed curve reproduces the SM triviality bound.}
\label{stability-fig}
\end{figure}
%%%%%%%%%%%%%%%%%%%%%%%%%%%%%%

The second constraint is potentially a stronger one.  Requiring that the 
quartic remain perturbative, $\lambda(\mu) \lsim 4 \pi$, we find
that the upper bound on the cutoff scale of the theory rapidly 
becomes small as the Higgs mass is increased.  We show this 
constraint as well as the meta-stability constraint in
Fig.~\ref{stability-fig}.
We find that for our choices of fourth--generation masses, 
the Yukawa interactions remain perturbative to slightly beyond the 
Higgs meta-stability/triviality bounds for all considered 
Higgs masses.  The ``chimney'' region, in which the effective theory 
of the Standard Model with $m_{H_{\rm SM}} \sim 200$ GeV remains valid to 
$\Mpl$, closes off. 
We find the maximal cutoff scale before new physics
of any kind enters occurs for Higgs masses in
the neighborhood of 300~GeV.  Much lower Higgs masses, 
in particular $m_H < 2 M_W$, imply other new physics must enter 
to prevent developing a deeper minimum away from the electroweak
breaking vacuum.  Nevertheless, we emphasize that this new physics 
can be weakly coupled below a TeV with little effect on Higgs physics
itself.

Conversely, to resolve the physics of the cutoff scale in the case
where the quartic (or the Yukawas) encounter a Landau pole 
undoubtedly requires physics directly connected to electroweak
symmetry breaking.  This new physics could be stronger-coupled 
supersymmetry, technicolor, topcolor, or a little Higgs construction.\bigskip

\section{Discussion}
\label{discussion-sec}

We have considered the constraints on a fourth generation 
and its effects on Higgs physics in the Standard Model.
If Nature does indeed have a fourth generation, 
it is amusing to speculate on the rich series of new phenomena 
expected at colliders now operating and about to begin.
The ordering of discoveries could proceed by Tevatron discovering 
the Higgs, with an unusually large production cross section, or in 
mass range that was previously thought to be undetectable in the 
Standard Model.
Subdominant decays of the Higgs may reveal a new sector.
Direct production of fourth generation neutrinos or leptons
may also be possible at Tevatron, but relies on a more
detailed understanding the background.  Once the LHC turns on,
the fourth generation quarks should be readily produced and found.
The Higgs can be found using the golden mode for a wide range 
of mass, and for most of this range, it will be found 
very quickly with a small integrated luminosity (due to the large
enhancement of the gluon fusion channel).  Given measures of the
cross section for Higgs production as well as branching ratios
of Higgs into subdominant modes, the LHC will be able to rapidly
verify that a fourth \emph{chiral} generation does indeed exist.

While our focus has been on the \emph{effects} of a fourth generation,
there is also the possibility that a fourth generation could 
alleviate or solve some of the pressing problems addressed by 
other models of new physics.  One amusing possibility is to employ
a variation of the mechanism of Ref.~\cite{Carena:2004ha} 
to revive electroweak 
baryogenesis in the (four-generation) Standard Model.  
Another possibility is to impose a parity symmetry to stabilize
the fourth generation lepton to serve as cold dark matter.
This is naively ruled out by direct detection, however there are
mechanisms \cite{Smith:2001hy,Belanger:2007dx} to avoid these bounds
by either splitting the neutrino eigenstates with a small Majorana mass
or otherwise invoking additional physics such as a $Z^\prime$
coupling to $U(1)_{B-L}$.  A detailed study of these issues is 
in progress and will be reported on elsewhere.

\begin{acknowledgments}

We thank D.E.~Kaplan, E.~Katz, M.~Schmaltz, and J.~Wacker for discussions 
and C.~Wagner for reminding us of the true and electroweak baryogenesis.
GDK, TP, and TT thank the Aspen Center of Physics for a stimulating
atmosphere where this work was begun.  GDK thanks SUPA
and Argonne National Laboratory for hospitality 
where part of this work was completed.
This work was supported in part by the Department of Energy 
under contracts DE-FG02-96ER40969 (GDK) and DE-AC02-06CH11357 (TT).

\end{acknowledgments}

%\bibliography{your bib file}

\begin{thebibliography}{99}
\bibitem{footnote0}
One could also speculate about five or more generations.
Optimizing over particle masses, we find that the contribution 
to the oblique electroweak parameters is $\Delta S \simeq \Delta T 
\gsim 0.3$ and thus outside the 95\% CL ellipse.
\bibitem{footnote2}
The Tevatron also has sensitivity to the charged and neutral leptons,
although there are several subtleties involving triggering 
(particularly when $p_T$ is small) and backgrounds.
We are not aware of robust bounds from Run II data
(independently of PMNS flavor mixing, for example).
However, it may be possible to eliminate some lepton mass 
parameter space for some or all flavor-mixings with a careful analysis.
\end{thebibliography}

\end{document}